\documentclass[10pt,draftcls,onecolumn]{IEEEtran}
\ifCLASSINFOpdf
\else
   \usepackage[dvips]{graphicx}
\fi
\usepackage{url}
\usepackage{graphicx}
\usepackage{amsmath}
\usepackage{comment}
\usepackage[utf8]{inputenc}
\usepackage[english]{babel}
\usepackage{color}
\usepackage{framed}
\usepackage{amsmath,amssymb,amsfonts}
\usepackage{graphicx}
\usepackage{upgreek}
\usepackage{textcomp}
\usepackage{url}
\usepackage[left=1.62cm,right=1.62cm,top=1.9cm]{geometry}
\usepackage{cite}

\usepackage{bbold}
\usepackage{ifpdf}
\usepackage{times}
\usepackage{layout}
\usepackage{float}
\usepackage{afterpage}
\usepackage{amsmath}
\usepackage{amstext}
\usepackage{amssymb,bm}
\usepackage{latexsym}
\usepackage{color}
\usepackage{flexisym}
\usepackage{kantlipsum}
\usepackage{graphicx}
\usepackage{amsmath}
\usepackage{amsthm}
\usepackage{graphicx}

\usepackage[caption=false,font=footnotesize]{subfig}

\usepackage{booktabs}
\usepackage{multicol}

\usepackage{enumitem}
\usepackage{mathtools}

\usepackage[switch]{lineno}
\usepackage[named]{algo}
\usepackage[noend]{algpseudocode}
\usepackage{algorithm}
\usepackage{amsmath}
\usepackage{algpseudocode}
\usepackage{amsthm}
\usepackage{dsfont}
\usepackage{algorithm}
\usepackage{thmtools}
\usepackage{amssymb}
\usepackage[dvipsnames]{xcolor}

\def\bGamma{\boldsymbol{\Gamma}}

\def\bOmega{\boldsymbol{\Omega}}

\def\mbb{\mathbf{b}}
\def\mbc{\mathbf{c}}

\def\mbe{\mathbf{e}}

\def\mbp{\mathbf{p}}

\def\mbr{\mathbf{r}}

\def\mbt{\mathbf{t}}
\def\mbu{\mathbf{u}}
\def\mbv{\mathbf{v}}

\def\mbx{\mathbf{x}}
\def\mby{\mathbf{y}}
\def\mbz{\mathbf{z}}

\def\mbA{\mathbf{A}}
\def\mbB{\mathbf{B}}
\def\mbC{\mathbf{C}}

\def\mbP{\mathbf{P}}

\def\mbR{\mathbf{R}}

\def\mbV{\mathbf{V}}

\def\mbX{\mathbf{X}}

\def\mbZ{\mathbf{Z}}

\newtheorem{proposition}{Proposition}
\newtheorem{lemma}{Lemma}

\theoremstyle{definition}

\algnewcommand\algorithmicinput{\textbf{Input:}}
\algnewcommand\Input{\item[\algorithmicinput]}
\algnewcommand\algorithmicoutput{\textbf{Output:}}
\algnewcommand\Output{\item[\algorithmicoutput]}
\algnewcommand\algorithmicinit{\textbf{Initialize:}}
\algnewcommand\Init{\item[\algorithmicinit]}

\makeatletter
\newcommand*{\rom}[1]{\expandafter\@slowromancap\romannumeral #1@}
\makeatother

\begin{document}
\title{Low-rank Matrix Sensing With Dithered One-Bit Quantization}

\author{\IEEEauthorblockN{Farhang Yeganegi$^{\star}$, Arian Eamaz$^{\star}$, and Mojtaba Soltanalian}\\
\IEEEauthorblockA{{University of Illinois Chicago, Chicago, IL 60607, USA}}
\thanks{This work was supported in part by the National Science Foundation Grant CCF-1704401.\\$\star$ The first two authors contributed equally to this work.}
}


\markboth{
}
{Shell \MakeLowercase{\textit{et al.}}: Bare Demo of IEEEtran.cls for IEEE Journals}

\maketitle
\thispagestyle{empty}
\pagestyle{empty}
\begin{abstract}
We explore the impact of coarse quantization on low-rank matrix sensing in the extreme scenario of \emph{dithered one-bit sampling}, where the high-resolution measurements are compared with random time-varying threshold levels. To recover the low-rank matrix of interest from the highly-quantized collected data, we offer an enhanced randomized Kaczmarz algorithm that efficiently solves the emerging highly-overdetermined feasibility problem. Additionally, we provide theoretical guarantees in terms of the convergence and sample size requirements. Our numerical results demonstrate the effectiveness of the proposed methodology.
\end{abstract}

\IEEEpeerreviewmaketitle

\section{Introduction}
\label{sec:intro}
The task of \emph{recovering a low-rank matrix from its linear measurements} plays a central role in computational science. The problem occurs in many areas of applied mathematics, such as signal processing \cite{davenport2016overview}, machine learning \cite{haeffele2014structured,nie2012low}, and computer vision \cite{tomasi1992shape}. Many approaches have been introduced in the literature to tackle this problem including singular value thresholding (SVT) \cite{cai2010singular}, singular value projection (SVP) \cite{jain2010guaranteed}, and low-rank matrix factorization \cite{chi2019nonconvex}.

Sampling signals at high data rates using high-resolution Analog-to-Digital Converters (ADCs) would significantly increase both manufacturing costs and power consumption. In multi-bit sampling scenarios, a very large number of quantization levels is necessary in order to represent the original continuous signal in with high accuracy, which in turn leads to a considerable reduction in sampling rate \cite{eamaz2021modified}. This attribute of multi-bit sampling is the key reason for the general emergence of underdetermined systems \cite{candes2013phaselift}. An alternative solution to such challenges is to deploy \emph{one-bit quantization} which is an extreme sampling scenario, where the signals are merely compared with given threshold levels at the ADCs, producing sign data ($\pm1$). This allows signal processing systems to sample at much higher rates while significantly reducing both cost and energy consumption compared to multi-bit ADC-based systems \cite{eamaz2021modified,sedighi2021performance}. Several applications abound of one-bit ADCs, such as multiple-input multiple-output wireless communications \cite{mezghani2018blind}, channel estimation \cite{li2017channel}, and array signal processing \cite{liu2017one}.

In classical one-bit sampling, signals are reconstructed by comparison with a fixed threshold, commonly set to zero. This method has limitations in accurately estimating signal parameters, especially when the input signal $\mathbf{x}$ is converted into one-bit data since the power information is lost. This occurs because the signs of $\mathbf{x}$ and $\eta\mathbf{x}$ are the same for $\eta>0$. The effectiveness of incorporating random thresholds (dithers) within the framework of one-bit quantization has been extensively established in various contexts \cite{dabeer2006signal}.
Extensive research has been dedicated to investigating the impact of thresholding (dithering) on the quantization process and its effect on quantization error \cite{gray1993dithered}. In the evaluation of quantizing systems, particularly in digital signal processing applications, the mean squared error (MSE) serves as a relevant parameter for assessing performance. Dithering provides the flexibility to control system performance by enabling a trade-off between accuracy and resolution. A notable feature of dithering is its capability to lower the average quantization error  \cite{carbone1997quantitative}.

The implementation of a dithered generator, specifically for generating Gaussian dithering in quantization systems within an ADC system, was described in \cite{robinson2019analog}. According to their work, the dither generator produces Gaussian dithering by generating a random analog noise signal. One method of achieving this is by utilizing a low-cost thermal noise diode, which introduces analog Gaussian dither. The intrinsic quantum mechanical properties of electron-hole pairing in such devices generate a truly random noise signal. In certain instances, the noise power generated by some dither generators may be relatively low. Therefore, optional processing circuitry can be employed alongside the dither generator, which may include components such as gain control circuitry for the analog noise signal, low-cost operational amplifiers, or similar elements, as required. Additionally, the implementation of multiple random dithering was demonstrated in \cite{ali2020background} for a 12-bit, 18-GS/s ADC.

The domain of one-bit low-rank matrix sensing has garnered considerable interest, yet it remains relatively unexplored. Several noteworthy papers, such as \cite{foucart2019recovering, foucart2019one}, have shed light on this intriguing subject. In \cite{foucart2019recovering}, the authors delve into the theoretical guarantees of one-bit sensing from low-rank matrices, employing two distinct algorithms: one based on hard singular value thresholding and the other utilizing semidefinite programming.
The initial investigation in this study considers a scenario \emph{without} thresholds and subsequently progresses to the Gaussian sampling matrix and Gaussian dithering scenario, where both adaptive and non-adaptive thresholds are employed. However, the presented recovery algorithms 
in this study require the availability of side information regarding the signal of interest, specifically an upper bound on the Frobenius norm of the input signal. This contradicts the fundamental advantage of using dithering in one-bit sensing, as it is generally understood to enable signal magnitude recovery without the need for extra information. In the notable work of \cite{foucart2019one}, the authors explore one-bit sensing for both low-rank and bisparse matrices. Notably, they achieve a significant milestone by deriving tailored theoretical guarantees for these types of matrices, providing a novel contribution to the field. Additionally, they investigate the impact of coarse quantization in thresholdless scenarios, focusing on recovering only the signal direction rather than its exact magnitude. It is worth mentioning that both of these influential papers exclusively focus on scenarios involving Gaussian sampling matrices.

The Randomized Kaczmarz Algorithm (RKA) \cite{leventhal2010randomized} is an iterative projection-based method for solving linear systems of equations and inequalities. It is particularly well-suited for highly overdetermined systems due to its simplicity. 
Each iteration projects onto the solution space corresponding to one row in the linear system, in a sequential regimen.  
In this paper, we address the problem of dithered one-bit low-rank matrix sensing by integrating the RKA with singular value projection (SVP).  We also provide the theoretical guarantees regarding the convergence rate of our proposed algorithm to the desired signal space. Section~\ref{sec2} begins with an introduction to the mathematical framework behind one-bit quantization, focusing on scenarios with multiple time-varying thresholds. Subsequently, we formulate the one-bit low-rank matrix sensing polyhedron, which we aim to solve via the integration of RKA with SVP. In Section~\ref{sec3}, we provide the uniform reconstruction guarantee for one-bit low-rank matrix sensing problem. Building upon this guarantee, we specify the conditions under which the RKA effectively converges to the desired signal space. The numerical results are presented in Section~\ref{sec4} to showcase the efficacy of our proposed method in one-bit low-rank matrix sensing. Finally, Section~\ref{sec5} concludes the paper.

\underline{\emph{Notation:}}
We use bold lowercase letters for vectors and bold uppercase letters for matrices. 
$\mathbb{R}$ represents the set of 
real numbers.
$(\cdot)^{\top}$ 
denotes the vector/matrix transpose.
$\operatorname{Tr}(.)$ denotes the trace of the matrix argument. $\left\langle \mbB_{1},\mbB_{2}\right\rangle=\operatorname{Tr}(\mbB_{1}^{\mathrm{H}}\mbB_{2})$ is the standard inner product between two
matrices. The nuclear norm of a matrix $\mbB\in \mathbb{R}^{n_{1}\times n_{2}}$ is denoted $\left\|\mbB\right\|_{\star}=\sum^{r}_{i=1}\sigma_{i}$ where $r$ and $\left\{\sigma_{i}\right\}$ are the rank and singular values of $\mbB$, respectively. The Frobenius norm of a matrix $\mbB$ is defined as $\|\mbB\|_{\mathrm{F}}=\sqrt{\sum^{n_{1}}_{t=1}\sum^{n_{2}}_{s=1}\left|b_{ts}\right|^{2}}$ where $\{b_{rs}\}$ are elements of $\mbB$. The $\ell_{k}$-norm of a vector $\mathbf{b}$ is defined as $\|\mbb\|^{k}_{k}=\sum_{i}|b|^{k}_{i}$. The Hadamard (element-wise) product of two matrices $\mbB_{1}$ and $\mbB_{2}$ is denoted as $\mbB_{1}\odot \mbB_{2}$. 
The vectorized form of a matrix $\mbB$ is written as $\operatorname{vec}(\mbB)$. 
For a given scalar $x$, we define $(x)^{+}$ as $\max\left\{x,0\right\}$. The set $[n]$ is defined as $[n]=\left\{1,\cdots,n\right\}$. $\operatorname{diag}\left\{\mathbf{b}\right\}$ denotes a diagonal matrix with $\{b_{i}\}$ as its diagonal elements. A ball with radius $\rho$ centered at a point $\mby\in\mathbb{R}^{n}$ is defined as $\mathcal{B}_\rho\left(\mby\right)=\left\{\mby_1\in\mathbb{R}^{n}|\left\|\mby-\mby_1\right\|_2\leq \rho\right\}$. 
$x \sim \mathcal{N}(\mu,\sigma^2)$ represents the normal distribution with mean $\mu$ and variance $\sigma^2$. 

\section{Problem Formulation}
\label{sec2}
Section~\ref{sec2_1} offers a succinct overview of the mathematical framework for one-bit quantization, specifically focusing on the incorporation of multiple time-varying threshold sequences. In Section~\ref{sec2_2}, the one-bit low-rank matrix sensing polyhedron is formed. To find the approximated solution within the desired signal space, we introduce the integration of RKA with SVP, called SVP-RKA, in Section~\ref{sec2_3}.

\subsection{One-Bit Quantization With Multiple Varying Thresholds}
\label{sec2_1}
Let $y_{k}=y(k\mathrm{T})$ denote the uniform samples of signal $y(t)$ with the sampling rate $1/\mathrm{T}$. 
In practice, the discrete-time samples occupy pre-determined quantized values. We denote the quantization operation on $y_{k}$ by the function $Q(\cdot)$. This yields the scalar quantized signal as $r_{k} = Q(y_{k})$. In one-bit quantization, compared to zero or constant thresholds, time-varying sampling thresholds yield a better recovery performance \cite{ameri2018one,eamaz2023covariance}. These thresholds may be chosen from any distribution. In the case of one-bit quantization with such time-varying sampling thresholds, we have
$r_{k} = \operatorname{sgn}\left(y_{k}-\tau_{k}\right)$. The information gathered through the one-bit sampling with time-varying thresholds presented here may be formulated in terms of an overdetermined linear system of inequalities. We have $r_{k}=+1$ when $y_{k}>\tau_{k}$ and $r_{k}=-1$ when $y_{k}<\tau_{k}$. Therefore, one can formulate the geometric location of the signal as
$r_{k}\left(y_{k}-\tau_{k}\right) \geq 0$. Collecting all the elements in the vectors as $\mathbf{y}=[y_{k}] \in \mathbb{R}^{n}$ and $\mathbf{r}=[r_{k}] \in \{-1,1\}^{n}$, we have $\mbr\odot\left(\mathbf{y}-\boldsymbol{\uptau}\right) \succeq \mathbf{0}$, or equivalently $\bOmega_{\mby}\mathbf{y}\succeq\mbr\odot \boldsymbol{\uptau}$,
where $\bOmega_{\mby} \triangleq \operatorname{diag}\left\{\mbr\right\}$. Denote the time-varying sampling threshold in $\ell$-th signal sequence by $\boldsymbol{\uptau}^{(\ell)}$, where $\ell\in [m]$. Then, we can write
\begin{equation}
\label{eq:7}
\begin{aligned}
\bOmega^{(\ell)}_{\mby} \mathbf{y} &\succeq \mathbf{r}^{(\ell)} \odot \boldsymbol{\uptau}^{(\ell)}, \quad \ell \in [m],
\end{aligned}
\end{equation}
where $\bOmega^{(\ell)}_{\mby}=\operatorname{diag}\left(\mbr^{(\ell)}\right)$. 
Denote the concatenation of all $m$ sign matrices as $\Tilde{\bOmega}_{\mby} =\left[\begin{array}{c|c|c}
\bOmega^{(1)}_{\mby}  &\cdots &\bOmega^{(m)}_{\mby} 
\end{array}\right]^{\top},\Tilde{\bOmega}_{\mby} \in \left\{-1,0,1\right\}^{m n\times n}$.
Rewrite the $m$ linear inequalities in \eqref{eq:7} as
\begin{equation}
\label{eq:8}
\Tilde{\bOmega}_{\mby}  \mathbf{y} \succeq \operatorname{vec}\left(\mbR_{\mby} \right)\odot \operatorname{vec}\left(\bGamma\right),
\end{equation}
where $\mathbf{R}_{\mby}$ and $\bGamma$ are matrices, whose columns are the sequences $\left\{\mathbf{r}^{(\ell)}\right\}_{\ell=1}^{m}$ and $\left\{\boldsymbol{\uptau}^{(\ell)}\right\}_{\ell=1}^{m}$, respectively.

Assuming a large number of samples --- a common situation in one-bit sampling scenarios --- hereafter we treat (\ref{eq:8}) as an overdetermined linear system of inequalities associated with the one-bit sensing scheme.
The inequality (\ref{eq:8}) can be recast as a polyhedron,
\begin{equation}
\label{eq:8n}
\begin{aligned}
\mathcal{P}_{\mby}=\left\{\mby^{\prime} \in \mathbb{R}^n \mid \tilde{\boldsymbol{\Omega}}_{\mby} \mby^{\prime} \succeq \operatorname{vec}\left(\mbR_{\mby}\right) \odot \operatorname{vec}(\boldsymbol{\Gamma})\right\} \subset \mathbb{R}^n,
\end{aligned}
\end{equation}
which we refer to as the \emph{one-bit polyhedron}. Generally, it can be assumed that the signal $\mbx\in\mathbb{R}^{d}$ is observed linearly through the sampling matrix $\mbA\in\mathbb{R}^{n\times d}$, creating the measurements as $\mby=\mbA\mbx$. Based on \eqref{eq:8}, the one-bit polyhedron for this type of problem is given by
\begin{equation}
\label{eq:80n}
\begin{aligned}
\mathcal{P}_{\mathbf{x}}=\left\{\mathbf{x}^{\prime}\in\mathbb{R}^{d} \mid \mbP_{\mby}  \mathbf{x}^{\prime} \succeq \operatorname{vec}\left(\mbR_{\mby} \right)\odot \operatorname{vec}\left(\bGamma\right)\right\}\subset \mathbb{R}^d,
\end{aligned}
\end{equation}
where $\mbP_{\mby} =\Tilde{\bOmega}_{\mby}\mbA$ or equivalently
\begin{equation}
\label{eq:90}
\mbP_{\mby} =\left[\begin{array}{c|c|c}
\mbA^{\top}\bOmega^{(1)}_{\mby}&\cdots &\mbA^{\top}\bOmega^{(m)}_{\mby} 
\end{array}\right]^{\top}, \quad \mbP_{\mby} \in\mathbb{R}^{m n\times d}.
\end{equation}

\subsection{One-Bit Low-Rank Matrix Sensing}
\label{sec2_2}
The problem of low-rank matrix sensing is formulated as:
\begin{equation}
\label{eq:1nnnnn}
\begin{aligned}
\text{find}\quad \mbX \in \Omega_{c} \quad
\text{subject to} \quad \mathcal{A}\left(\mbX\right)=\mathbf{y},~ 
\operatorname{rank}\left(\mbX\right)\leq r,
\end{aligned}
\end{equation}
where $\mbX\in \mathbb{R}^{n_{1}\times n_{2}}$ is the matrix of unknowns, $\mathbf{y}\in \mathbb{R}^{n}$ is the measurement vector, and $\mathcal{A}$ is a linear transformation such that $\mathcal{A}:\mathbb{R}^{n_1\times n_2}\mapsto\mathbb{R^{n}}$.
In general, $\Omega_{c}$ can be chosen such as the set of semi-definite matrices, symmetric matrices, upper or lower triangle matrices, Hessenberg matrices and a specific constraint on the matrix elements $\left\|\mbX\right\|_{\infty}\leq \alpha$ or on its eigenvalues, i.e., $\lambda_{i}\leq \epsilon$ where $\left\{\lambda_{i}\right\}$ are eigenvalues of $\mbX$ \cite{davenport2016overview,candes2015phase,van1996matrix}. The problem (\ref{eq:1nnnnn}) can be rewritten as an optimization problem:
\begin{equation}
\label{eq:1nnnnnn}
\begin{aligned}
\underset{\mbX \in \Omega_{c}}{\textrm{minimize}}\quad \operatorname{rank}\left(\mbX\right) \quad
\text{subject to} \quad \mathcal{A}\left(\mbX\right)=\mathbf{y}.
\end{aligned}
\end{equation}
This problem is known to be NP-hard, whose solution is difficult to approximate \cite{meka2008rank,recht2011null}. Many approaches have been introduced in the literature to tackle this problem (or its relaxed version) including SVT \cite{cai2010singular}, SVP \cite{jain2010guaranteed}, and low-rank matrix factorization \cite{chi2019nonconvex,jain2010guaranteed}. In low-rank matrix sensing, the linear operator $\mathcal{A}\left(\mbX\right)$ is obtained as \cite{chi2019nonconvex},
\begin{equation}
\label{Stefanie_2}
\mathcal{A}\left(\mbX\right)=\frac{1}{\sqrt{n}}\left[\operatorname{Tr}\left(\mbA^{\top}_1\mbX\right)\cdots\operatorname{Tr}\left(\mbA^{\top}_n\mbX\right)\right]^{\top},
\end{equation}
where $\mbA_j\in\mathbb{R}^{n_1\times n_2}$ is the $j$-th sensing matrix. Following the formulation provided in Section~\ref{sec2_1}, the one-bit polyhedron for the low-rank matrix sensing is given by
\begin{equation}
\label{kumar}
\mathcal{P}^{(M)}=\left\{\mbX^{\prime}\in\mathbb{R}^{n_1\times n_2} \mid r^{(\ell)}_{j}\operatorname{Tr}\left(\mbA^{\top}_j\mbX^{\prime}\right)\geq r^{(\ell)}_{j}\tau^{(\ell)}_{j}\right\}\subset \mathbb{R}^{n_1\times n_2},
\end{equation} 
for all $j\in[n],\ell\in[m]$. In order to obtain the solution within a reduced number of samples in the polyhedron $\mathcal{P}^{(M)}$ defined in \eqref{kumar}, we impose a rank constraint, $\operatorname{rank}(\mbX)\leq r$, to shrink the entire space, as shown by the following polyhedron:
\begin{equation}
\label{Stefanie_5}
\mathcal{P}^{(M)}_1=\left\{\mbX^{\prime}\in\mathcal{P}^{(M)} \mid \operatorname{rank}\left(\mbX^{\prime}\right)\leq r\right\}\subset \mathbb{R}^{n_1\times n_2}.
\end{equation}
In Section~\ref{sec3_1}, we will provide the required number of one-bit samples to achieve a uniform reconstruction result in the polyhedron \eqref{Stefanie_5} with high probability.

\subsection{SVP-RKA}
\label{sec2_3}
The RKA serves as a \emph{sub-conjugate gradient method} for solving linear feasibility problems of the form $\mbC\mathbf{x}\succeq\mathbf{b}$, where $\mbC$ is a ${m\times n}$ matrix with $m>n$ \cite{leventhal2010randomized}. Conjugate-gradient methods immediately turn the mentioned inequality to an equality in the following form $\left(\mbb-\mbC\mathbf{x}\right)^{+}=0$,
and then, approach the solution by the same process as used for systems of equations. The projection coefficient $\beta_{i}$ of the RKA is
\begin{equation}
\label{eq:22}
\beta_{i}= \begin{cases}
\left(b_{j}-\langle\mathbf{c}_{j},\mathbf{x}_{i}\rangle\right)^{+} & \left(j \in \mathcal{I}_{\geq}\right), \\ b_{j}-\langle\mathbf{c}_{j},\mathbf{x}_{i}\rangle & \left(j \in \mathcal{I}_{=}\right),
\end{cases}
\end{equation}
where the disjoint index sets $\mathcal{I}_{\geq}$ and $\mathcal{I}_{=}$ partition $[m]$ and $\{\mathbf{c}_{j}\}$ are the rows of $\mathbf{C}$.
Also, the unknown column vector $\mathbf{x}$ is iteratively updated as $\mathbf{x}_{i+1}=\mathbf{x}_{i}+\frac{\beta_{i}}{\left\|\mbc_{j}\right\|^{2}_{2}} \mbc^{\star}_{j}$,
where at each iteration $i$, the index $j$ is drawn from the set $[m]$ independently at random following the distribution
$\operatorname{Pr}\{j=k\}=\frac{\left\|\mbc_{k}\right\|^{2}_{2}}{\|\mbC\|_{\mathrm{F}}^{2}}$.
Assuming that the linear system is consistent with nonempty feasible set $\mathcal{P}_{\mbx}$ created by the intersection
of hyperplanes around the desired point $\mbx$,
RKA converges linearly in expectation to the solution $\widehat{\mbx}\in\mathcal{P}_{\mbx}$\cite{leventhal2010randomized}:
\begin{equation}
\label{eq:15}
\mathbb{E}\left\{\hbar\left(\mathbf{x}_{i},\widehat{\mbx}\right)\right\} \leq \left(1-q_{_{\text{RKA}}}\right)^{i}~ \hbar\left(\mathbf{x}_{0},\widehat{\mbx}\right),
\end{equation}
where $\hbar\left(\mathbf{x}_{i},\widehat{\mbx}\right)=\left\|\mathbf{x}_{i}-\widehat{\mbx}\right\|_{2}^{2}$, is the euclidean distance between two points in the space, 
$i$ is the number of required iterations for RKA, and $q_{_{\text{RKA}}} \in \left(0,1\right)$ is given by $q_{_{\text{RKA}}}=\frac{1}{\kappa^{2}\left(\mbC\right)}$,
with $\kappa\left(\mbC\right)=\|\mbC\|_{\mathrm{F}}\|\mbC^{\dagger}\|_{2}$ denoting the scaled condition number of a matrix $\mbC$. The robustness of the RKA
against noise has been demonstrated in \cite{needell2010randomized}. Furthermore, the authors of \cite{huang2022linear} specifically explored the performance of the RKA 
in the presence of \emph{Gaussian} and \emph{Poisson} noise, highlighting its robustness even when dealing with Poisson noisy measurements.

The SVP was introduced as a solution to the general affine rank minimization problem (ARMP). In \cite{jain2010guaranteed}, it was demonstrated that the SVP can effectively recover the minimum rank solution even in the presence of noise and when the affine constraints satisfy the restricted isometry property (RIP). Moreover, some theoretical guarantees for this approach were also established. This method utilizes the operator $P_{r}$ to modify the gradient descent process at each iteration. The operator $P_{r}$ calculates the $r$ largest singular values of a matrix and subsequently rewrites its singular value decomposition (SVD) based on these $r$ singular values and their corresponding singular vectors. To find an approximated solution within the polyhedron \eqref{Stefanie_5}, we propose the integration of SVP into each iteration of RKA which results in the following update process:
\begin{equation}
\label{St_20}
\left\{\begin{array}{l}
\mbZ_{i+1}=\mbX_i+\frac{\left(r^{(\ell)}_j\tau^{(\ell)}_j-r^{(\ell)}_j\operatorname{Tr}\left(\mbA^{\top}_j\mbX_i\right) \right)^{+}}{\left\|\mbA_{j}\right\|^{2}_{\mathrm{F}}} \mbA_{j},\\
\mbX_{i+1}=P_{r}\left(\mbZ_{i+1}\right).
\end{array}\right.
\end{equation}
Within Section~\ref{sec3_2}, we provide the convergence rate for the update process of SVP-RKA, as detailed in \eqref{St_20}.

\section{Theoretical Guarantees}
\label{sec3}
Section~\ref{sec3_1} provides the uniform reconstruction guarantee for approximating a low-rank matrix within the polyhedron \eqref{Stefanie_5}. Subsequently, Section~\ref{sec3_2} outlines the convergence rate of our proposed algorithm, SVP-RKA, toward the desired signal space.

\subsection{Uniform Reconstruction Guarantee}
\label{sec3_1}
In order to establish the minimum number of one-bit samples $m^{\prime}=mn$ such that the uniform reconstruction guarantee can be achieved in the polyhedron \eqref{Stefanie_5}, we only consider one sequence of time-varying sampling thresholds, $m=1$, which results in $m^{\prime}=n$. This assumption is reasonable because our objective is to leverage the polyhedron \eqref{Stefanie_5} to achieve an optimal low-rank solution with a limited number of one-bit samples. To present our theoretical guarantee, we additionally consider the Frobenius norm constraint, $\|\mbX\|_{\mathrm{F}}\leq 1$, in the polyhedron \eqref{Stefanie_5} leading to the definition of the set
\begin{equation}
\label{kumkum}
\bar{\mathcal{K}}_{n_1,n_2,r}=\left\{\mbX^{\prime}\in\mathbb{R}^{n_1\times n_2}\mid\operatorname{rank}(\mbX^{\prime})\leq r,\|\mbX^{\prime}\|_{\mathrm{F}}\leq 1\right\}.
\end{equation}
The following proposition outlines the necessary number of one-bit samples $n$ for recovering the optimal low-rank matrix $\mbX$ from the polyhedron \eqref{Stefanie_5}:
\begin{proposition}[\textit{Random Hyperplane Tessellations of $\bar{\mathcal{K}}_{n_1,n_2,r}$}]
\label{kvm_20}
Let each element of $\mbA_j\in\mathbb{R}^{n_1\times n_2}$ and each $\tau_j$ for $j\in[n]$ be independently drawn from the standard normal distribution. If $n\geq C\delta^{-4}(n_1+n_2)r$, then with probability at least $1-2e^{-c\delta^4n}$, all $\mbX,\mbX^{\prime}\in\bar{\mathcal{K}}_{n_1,n_2,r}$ with
\begin{equation}
\label{kvm_22}
\operatorname{sgn}\left(\operatorname{Tr}\left(\mbA_j^{\top}\mbX\right)-\tau_j\right)=\operatorname{sgn}\left(\operatorname{Tr}\left(\mbA_j^{\top}\mbX^{\prime}\right)-\tau_j\right),~j\in[n],
\end{equation}
satisfy $\operatorname{vec}(\mbX^{\prime})\in\mathcal{B}_{\frac{\delta}{4}}(\operatorname{vec}(\mbX))$.
The positive constants $c$ and $C$ are absolute constants.
\end{proposition}
\begin{IEEEproof}
We will prove this proposition by considering the random hyperplane tessellations theorem for the set $\Tilde{\mathcal{K}}_{n_1,n_2,r}=\left\{\mbX^{\prime}\in\mathbb{R}^{n_1\times n_2}\mid\operatorname{rank}(\mbX^{\prime})\leq r,\|\mbX^{\prime}\|_{\mathrm{F}}= 1\right\}$ in the ditherless scenario. As provided in \cite[Theorem~3.1]{plan2014dimension} for the Gaussian sensing matrix, if $n\geq C\delta^{-4}(n_1+n_2)r$, then with probability at least $1-2e^{-c\delta^4n}$, all $\mbX,\mbX^{\prime}\in\Tilde{\mathcal{K}}_{n_1,n_2,r}$ with consistent reconstruction property, i.e., $\operatorname{sgn}\left(\operatorname{Tr}\left(\mbA_j^{\top}\mbX\right)\right)=\operatorname{sgn}\left(\operatorname{Tr}\left(\mbA_j^{\top}\mbX^{\prime}\right)\right)$ for all $j\in[n]$, satisfy $\|\mbX-\mbX^{\prime}\|_{\mathrm{F}}\leq \frac{\delta}{8}$. It is obvious that Proposition~\ref{kvm_20} is a translation of this theorem in the case of time-varying sampling thresholds for a tessellation of $\bar{\mathcal{K}}_{n_1,n_2,r}$ defined in \eqref{kumkum}. Define $\Tilde{\mbA}_{j}\in\mathbb{R}^{(n_1+1)\times (n_2+1)}$ as
\begin{equation}
\label{B2_4}
\Tilde{\mbA}_j=
\begin{bmatrix}
\mbA_j & \mbu_j \\
\mbv_j^{\top} & -\tau_j 
\end{bmatrix},~j\in[n],
\end{equation}
where each element of $\mbu_j\in\mathbb{R}^{n_1}$ and $\mbv_j\in\mathbb{R}^{n_2}$ is drawn independently from the standard normal distribution. Similarly, define $\Tilde{\mbX}\in\mathbb{R}^{(n_1+1)\times (n_2+1)}$ as
\begin{equation}
\label{B2_5}
\Tilde{\mbX}=\begin{bmatrix}
\mbX & \mathbf{0}_{n_1} \\
\mathbf{0}_{n_2}^{\top} & 1 
\end{bmatrix},
\end{equation}
where $\mathbf{0}_{n_1}$ represents the zero vector in an $n_1$-dimensional space, while $\mathbf{0}_{n_2}$ represents the zero vector in an $n_2$-dimensional space. Based on the definitions of $\Tilde{\mbA}_{j}$ and $\Tilde{\mbX}$, for any matrix $\mbX\in\bar{\mathcal{K}}_{n_1,n_2,r}$, we notice that 
\begin{equation}
\label{B2_6}
\operatorname{sgn}\left(\operatorname{Tr}\left(\mbA_j^{\top}\mbX\right)-\tau_j\right)=\operatorname{sgn}\left(\operatorname{Tr}\left(\Tilde{\mbA}_j^{\top}\Tilde{\mbX}\right)\right),~j\in[n].
\end{equation}
Thus, we have moved to the ditherless setup in the augmented space presented by $\Tilde{\mbA}_j$ and $\Tilde{\mbX}$. Note that $\Tilde{\mbX}$ has the rank at most $r+1$. Since $\|\Tilde{\mbX}\|_{\mathrm{F}}\geq 1$, we may apply the result of ditherless scenario after projecting on the space $\|\mbX\|_{\mathrm{F}}=1$ to derive
\begin{equation}
\label{B2_7}
\left\|\frac{\Tilde{\mbX}}{\|\Tilde{\mbX}\|_{\mathrm{F}}}-\frac{\Tilde{\mbX}^{\prime}}{\|\Tilde{\mbX}^{\prime}\|_{\mathrm{F}}}\right\|_{\mathrm{F}}\leq \frac{\delta}{8}.
\end{equation}
Based on \cite[Lemma~7]{foucart2019recovering}, one can obtain the corresponding bound on $\|\mbX-\mbX^{\prime}\|_{\mathrm{F}}$ as $\|\mbX-\mbX^{\prime}\|_{\mathrm{F}}\leq 2\left\|\frac{\Tilde{\mbX}}{\|\Tilde{\mbX}\|_{\mathrm{F}}}-\frac{\Tilde{\mbX}^{\prime}}{\|\Tilde{\mbX}^{\prime}\|_{\mathrm{F}}}\right\|_{\mathrm{F}}\leq \frac{\delta}{4}$,
which proves the proposition.
\end{IEEEproof}
Note that Proposition~\ref{kvm_20} presents a uniform reconstruction result, indicating that with high probability, all low-rank matrices can be reconstructed. This differs from a nonuniform result, where each low-rank matrix is individually reconstructed with high probability.

\subsection{Convergence Guarantee for SVP-RKA}
\label{sec3_2}
As formulated in \eqref{eq:15}, it is evident that the RKA converges linearly in expectation to the solution $\widehat{\mbx}$ within the feasible set $\mathcal{P}_{\mbx}$—a set formed by the intersection of one-bit hyperplanes defined in \eqref{eq:80n}. In this section, we aim to extend this result to the context of SVP-RKA, ensuring that the approximated solution falls inside the desired signal space, i.e., $\operatorname{vec}(\widehat{\mbX})\in\mathcal{B}_{\rho}(\operatorname{vec}(\mbX))$ for a small constant value of $\rho$. The subsequent lemma presents the convergence guarantee for SVP-RKA under these conditions:
\begin{lemma}
\label{Ea}
The update process of SVP-RKA presented in \eqref{St_20} converges linearly in expectation to a ball centered at the original signal
$\mathcal{B}_{\rho}\left(\operatorname{vec}\left(\mbX\right)\right)$ 
with the number of samples satisfying Proposition~\ref{kvm_20} and probability exceeding $1-2e^{-c\rho^4n}$, as follows:
\begin{equation}
\label{kvm_19}
\begin{aligned}
\mathbb{E}\left\{\left\|\mbX_{i}-\mbX\right\|_{\mathrm{F}}\right\}\leq\left(1-\frac{1}{\kappa^{2}\left(\mbV\right)}\right)^{\frac{i}{2}} \left\|\mbX_{0}-\widehat{\mbX}\right\|_{\mathrm{F}}+\rho,
\end{aligned}   
\end{equation}
where $\widehat{\mbX}\in \mathcal{P}^{(M)}_{1}$, and $\mbV$ is the matrix with vectorized sensing matrices $\left\{\operatorname{vec}\left(\mbA_j\right)\right\}^{n}_{j=1}$ as its rows. 
\end{lemma}
\begin{IEEEproof}
Denote $\mbX_i=P_r(\mbZ_i)$.
For simplicity of notations, assume $\mbx_i=\operatorname{vec}(\mbX_i)$, $\widehat{\mbx}=\operatorname{vec}(\widehat{\mbX})$, and $\mbx=\operatorname{vec}(\mbX)$. Also, denote $\mbP_{\mby}$ in \eqref{eq:90} by $\mbP$. By defining $\mbe_i=\mbx_i-\widehat{\mbx}$, we can write
\begin{equation}
\begin{aligned}
\label{kvm_3}
\left\|\mbx_i-\mbx\right\|_{2} &= \left\|\mbx_i-\widehat{\mbx}+\widehat{\mbx}-\mbx\right\|_{2}\\&\leq\left\|\mbe_i\right\|_{2}+\left\|\mbx-\widehat{\mbx}\right\|_{2}\\&=\left\|P_r\left(\mbZ_i\right)-P_r\left(\widehat{\mbX}\right)\right\|_{\mathrm{F}}+\left\|\mbx-\widehat{\mbx}\right\|_{2},
\end{aligned}   
\end{equation}
where we have utilized the fact that $\widehat{\mbX}\in \mathcal{P}^{(M)}_{1}$.
Consider an operator function $\mathcal{G}_f$ applied to a matrix $\mbX$ with rank $r$ as follows:
\begin{equation}
\label{kvm_34}
\mathcal{G}_f\left(\mbX\right)=\sum^{r}_{k=1} f\left(\sigma_k\right)\mbu_k\mbv^{\mathrm{H}}_k,
\end{equation}
where $\left\{\sigma_k, \mbu_k, \mbv_k\right\}^{r}_{k=1}$ are singular values of $\mbX$ and its corresponding singular 
vectors, and $f$ is a $L$-Lipschitz continuous projector function. 
As comprehensively discussed in \cite{carlsson2021lipschitz} and \cite[Theorem~4.2]{andersson2016operator}, the following relation holds for two matrices $\mbX_1$ and $\mbX_2$ belonging to the Hilbert space $\mathcal{H}$:
\begin{equation}
\label{Lip1}
\left\|\mathcal{G}_f\left(\mbX_1\right)-\mathcal{G}_f\left(\mbX_2\right)\right\|_{\mathrm{F}} \leq L \left\|\mbX_1-\mbX_2\right\|_{\mathrm{F}}.
\end{equation}
In 
SVP-RKA, $f$ 
is an operator that only chooses the $r$-largest singular values. 
It is straightforward to verify that such $f$ satisfies \eqref{Lip1} with $L=1$. Therefore, one can conclude
\begin{equation}
\label{Lip2}
\left\|P_r\left(\mbX_1\right)-P_r\left(\mbX_2\right)\right\|_{\mathrm{F}} \leq \left\|\mbX_1-\mbX_2\right\|_{\mathrm{F}},~\forall\mbX_1, \mbX_2 \in \mathcal{H}.
\end{equation}
Combining this result with \eqref{kvm_3} leads to
\begin{equation}
\label{kvm_4}
\left\|\mbx_i-\mbx\right\|_{2}\leq \left\|\mbz_i-\widehat{\mbx}\right\|_{2}+ \left\|\mbx-\widehat{\mbx}\right\|_{2},
\end{equation}
where $\mbz_i=\operatorname{vec}(\mbZ_i)$. Define $t^{(\ell)}_{j}=r^{(\ell)}_j\tau^{(\ell)}_j$ and $\mbp^{(\ell)}_j=r^{(\ell)}_j\mbv_j$, with $\mbv_j$ denoting the $j$-th row of the matrix $\mbV$. For the term $\left\|\mbz_i-\widehat{\mbx}\right\|_{2}$ in \eqref{kvm_4}, we have 
\begin{equation}
\begin{aligned}
\label{kvm_31}
\left\|\mbz_{i}-\widehat{\mbx}\right\|^2_2
&=\left\|\mbe_{i-1}+\frac{\left(t^{(\ell)}_j-\langle\mbp^{(\ell)}_j,\mbx_{i-1}\rangle\right)^+}{\left\|\mbp^{(\ell)}_j\right\|^2_2}  \mbp^{(\ell)}_j\right\|^2_2\\&=\left\|\mbe_{i-1}\right\|^2_2+\frac{\left(\left(t^{(\ell)}_j-\langle\mbp^{(\ell)}_j,\mbx_{i-1}\rangle\right)^+\right)^2}{\left\|\mbp^{(\ell)}_j\right\|^2_2}+\frac{2\left(t^{(\ell)}_j-\langle\mbp^{(\ell)}_j,\mbx_{i-1}\rangle\right)^+\langle\mbp^{(\ell)}_j,\mbe_{i-1}\rangle}{\left\|\mbp^{(\ell)}_j\right\|^2_2}.
\end{aligned}
\end{equation} 
Since $\langle\mbp^{(\ell)}_j,\widehat{\mbx}\rangle\geq t^{(\ell)}_j$, we have $\langle\mbp^{(\ell)}_j,\mbe_{i-1}\rangle=\langle\mbp^{(\ell)}_j,\mbx_{i-1}-\widehat{\mbx}\rangle\leq\langle\mbp^{(\ell)}_j,\mbx_{i-1}\rangle-t^{(\ell)}_j$. Therefore, one can rewrite \eqref{kvm_31} as
\begin{equation}
\begin{aligned}
\label{kvm_32}
\left\|\mbz_{i}-\widehat{\mbx}\right\|^2_2&\leq\left\|\mbe_{i-1}\right\|^2_2+\frac{\left(\left(t^{(\ell)}_j-\langle\mbp^{(\ell)}_j,\mbx_{i-1}\rangle\right)^+\right)^2}{\left\|\mbp^{(\ell)}_j\right\|^2_2}+\frac{2\left(t^{(\ell)}_j-\langle\mbp^{(\ell)}_j,\mbx_{i-1}\rangle\right)^+\langle\mbp^{(\ell)}_j,\mbe_{i-1}\rangle}{\left\|\mbp^{(\ell)}_j\right\|^2_2}\\
&\leq\left\|\mbe_{i-1}\right\|^2_2+\frac{\left(\left(t^{(\ell)}_j-\langle\mbp^{(\ell)}_j,\mbx_{i-1}\rangle\right)^+\right)^2}{\left\|\mbp^{(\ell)}_j\right\|^2_2}-\frac{2\left(t^{(\ell)}_j-\langle\mbp^{(\ell)}_j,\mbx_{i-1}\rangle\right)^+\left(t^{(\ell)}_j-\langle\mbp^{(\ell)}_j,\mbx_{i-1}\rangle\right)}{\left\|\mbp^{(\ell)}_j\right\|^2_2}\\
&=\left\|\mbe_{i-1}\right\|^2_2-\frac{\left(\left(t^{(\ell)}_j-\langle\mbp^{(\ell)}_j,\mbx_{i-1}\rangle\right)^+\right)^2}{\left\|\mbp^{(\ell)}_j\right\|^2_2}.
\end{aligned}
\end{equation}
Define $\mbt=\left[\begin{array}{c|c|c}
\mbt_1^{\top} &\cdots &\mbt_m^{\top}
\end{array}\right]^{\top}$, where $\mbt_{\ell}=\left[t_j^{(\ell)}\right]_{j=1}^{n}$ for $\ell\in[m]$. Taking the expectation from both sides of \eqref{kvm_32} results in
\begin{equation}
\begin{aligned}
\label{kvm_33}
\mathbb{E}\left\{\left\|\mbz_{i}-\widehat{\mbx}\right\|^2_2\right\}\leq\left\|\mbe_{i-1}\right\|^2_2-\frac{\left\|\left(\mbt-\mbP\mbx_{i-1}\right)^+\right\|^2_2}{\left\|\mbP\right\|^2_{\mathrm{F}}},
\end{aligned}
\end{equation}
where based on the Hoffman bound \cite[Theorem~4.2]{leventhal2010randomized}, we have
\begin{equation}
\begin{aligned}
\label{amoo_arian}
\mathbb{E}\left\{\left\|\mbz_{i}-\widehat{\mbx}\right\|^2_2\right\} \leq\left(1-\frac{1}{\kappa^{2}\left(\mbP\right)}\right) \left\|\mbe_{i-1}\right\|^2_2.
\end{aligned}
\end{equation}
It can be simply shown that $\kappa\left(\mbV\right)=\kappa\left(\mbP\right)$ which together with \eqref{amoo_arian} after $i$ iterations leads to
\begin{equation}
\label{Conv}
\mathbb{E}\left\{\left\|\mbz_{i}-\widehat{\mbx}\right\|_2\right\}\leq\left(1-\frac{1}{\kappa^{2}\left(\mbV\right)}\right)^{\frac{i}{2}} \left\|\mbe_0\right\|_2.
\end{equation}
Combining this result with \eqref{kvm_4} and 
utilizing the guarantees provided in Proposition~\ref{kvm_20} complete the proof.
\end{IEEEproof}
\begin{figure}[t]
	\centering
	\includegraphics[width=0.5\columnwidth]{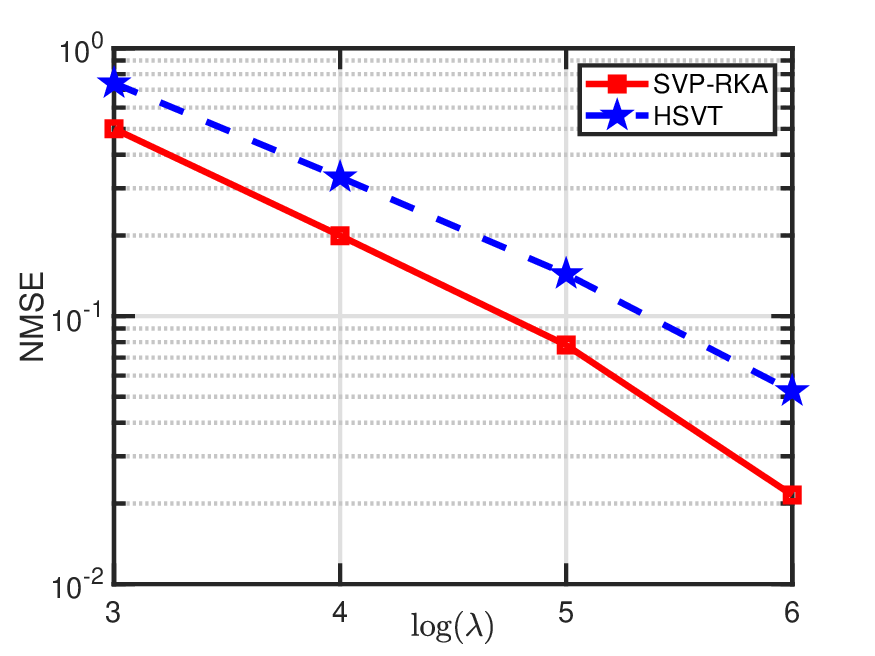} \vspace{-.2cm}
	\caption{Comparison between the recovery performance of SVP-RKA and HSVT algorithm over different values of oversampling factor $\lambda$.
	}
    \vspace{-10pt}
    \label{figure_1}
\end{figure}
\section{Numerical Results}
\label{sec4}
This section presents a set of numerical evaluations to assess the efficacy of the proposed algorithm. All presented results are averaged over 1000 experiments.
We generated a collection of sampling matrices $\{\mbA_j\}_{j=1}^{n}$, where each entry is independently sampled from a standard normal distribution. The desired matrix $\mbX\in\mathbb{R}^{30\times 30}$ was generated with $\operatorname{rank}(\mbX)=2$. Define the oversampling factor as $\lambda=\frac{n}{n_1r}=\frac{n}{60}$. In our experiments, we have set 
$\log(\lambda)\in\{3,4,5,6\}$. The number of time-varying sampling threshold sequences was fixed at $m=1$. Accordingly, we have generated sequences of time-varying sampling thresholds as $\left\{\boldsymbol{\uptau}^{(\ell)}\sim \mathcal{N}\left(\mathbf{0},\frac{\beta_{\mby}^2}{9}\mathbf{I}\right)\right\}_{\ell=1}^{m}$, where $\beta_{\mby}$ denotes the dynamic range of the noisy high-resolution measurements $\mby$. Fig.~\ref{figure_1} compares the recovery performance of SVP-RKA with hard singular value thresholding (HSVT) algorithm \cite{foucart2019recovering}. As can be observed, SVP-RKA outperforms HSVT over different values of the oversampling factor.

\section{Summary}
\label{sec5}
This study establishes a uniform reconstruction framework for the problem of one-bit low-rank matrix sensing under time-varying sampling thresholds. We studied the convergence of our proposed approach, SVP-RKA, to the signal of interest. Numerical experiments confirmed that SVP-RKA outperforms its state-of-the-art counterpart, namely HSVT.

\bibliographystyle{IEEEtran}
\bibliography{refs}

\begin{thebibliography}{10}
\providecommand{\url}[1]{#1}
\csname url@samestyle\endcsname
\providecommand{\newblock}{\relax}
\providecommand{\bibinfo}[2]{#2}
\providecommand{\BIBentrySTDinterwordspacing}{\spaceskip=0pt\relax}
\providecommand{\BIBentryALTinterwordstretchfactor}{4}
\providecommand{\BIBentryALTinterwordspacing}{\spaceskip=\fontdimen2\font plus
\BIBentryALTinterwordstretchfactor\fontdimen3\font minus
  \fontdimen4\font\relax}
\providecommand{\BIBforeignlanguage}[2]{{%
\expandafter\ifx\csname l@#1\endcsname\relax
\typeout{** WARNING: IEEEtran.bst: No hyphenation pattern has been}%
\typeout{** loaded for the language `#1'. Using the pattern for}%
\typeout{** the default language instead.}%
\else
\language=\csname l@#1\endcsname
\fi
#2}}
\providecommand{\BIBdecl}{\relax}
\BIBdecl

\bibitem{davenport2016overview}
M.~A. Davenport and J.~Romberg, ``An overview of low-rank matrix recovery from
  incomplete observations,'' \emph{IEEE Journal of Selected Topics in Signal
  Processing}, vol.~10, no.~4, pp. 608--622, 2016.

\bibitem{haeffele2014structured}
B.~Haeffele, E.~Young, and R.~Vidal, ``Structured low-rank matrix
  factorization: Optimality, algorithm, and applications to image processing,''
  in \emph{International conference on machine learning}.\hskip 1em plus 0.5em
  minus 0.4em\relax PMLR, 2014, pp. 2007--2015.

\bibitem{nie2012low}
F.~Nie, H.~Huang, and C.~Ding, ``Low-rank matrix recovery via efficient
  schatten p-norm minimization,'' in \emph{Twenty-sixth AAAI conference on
  artificial intelligence}, 2012.

\bibitem{tomasi1992shape}
C.~Tomasi and T.~Kanade, ``Shape and motion from image streams under
  orthography: a factorization method,'' \emph{International journal of
  computer vision}, vol.~9, no.~2, pp. 137--154, 1992.

\bibitem{cai2010singular}
J.~Cai, E.~Cand{\`e}s, and Z.~Shen, ``A singular value thresholding algorithm
  for matrix completion,'' \emph{SIAM Journal on optimization}, vol.~20, no.~4,
  pp. 1956--1982, 2010.

\bibitem{jain2010guaranteed}
P.~Jain, R.~Meka, and I.~Dhillon, ``Guaranteed rank minimization via singular
  value projection,'' \emph{Advances in Neural Information Processing Systems},
  vol.~23, 2010.

\bibitem{chi2019nonconvex}
Y.~Chi, Y.~Lu, and Y.~Chen, ``Nonconvex optimization meets low-rank matrix
  factorization: An overview,'' \emph{IEEE Transactions on Signal Processing},
  vol.~67, no.~20, pp. 5239--5269, 2019.

\bibitem{eamaz2021modified}
A.~Eamaz, F.~Yeganegi, and M.~Soltanalian, ``Modified arcsine law for one-bit
  sampled stationary signals with time-varying thresholds,'' in \emph{ICASSP
  2021-2021 IEEE International Conference on Acoustics, Speech and Signal
  Processing (ICASSP)}.\hskip 1em plus 0.5em minus 0.4em\relax IEEE, 2021, pp.
  5459--5463.

\bibitem{candes2013phaselift}
E.~J. Candes, T.~Strohmer, and V.~Voroninski, ``{PhaseLift}: Exact and stable
  signal recovery from magnitude measurements via convex programming,''
  \emph{Communications on Pure and Applied Mathematics}, vol.~66, no.~8, pp.
  1241--1274, 2013.

\bibitem{sedighi2021performance}
S.~Sedighi, M.~R.~B. Shankar, M.~Soltanalian, and B.~Ottersten, ``On the
  performance of one-bit {DoA} estimation via sparse linear arrays,''
  \emph{IEEE Transactions on Signal Processing}, vol.~69, pp. 6165--6182, 2021.

\bibitem{mezghani2018blind}
A.~Mezghani and A.~L. Swindlehurst, ``Blind estimation of sparse broadband
  massive {MIMO} channels with ideal and one-bit {ADCs},'' \emph{IEEE
  Transactions on Signal Processing}, vol.~66, no.~11, pp. 2972--2983, 2018.

\bibitem{li2017channel}
Y.~Li, C.~Tao, G.~Seco-Granados, A.~Mezghani, A.~Swindlehurst, and L.~Liu,
  ``Channel estimation and performance analysis of one-bit massive {MIMO}
  systems,'' \emph{IEEE Transactions on Signal Processing}, vol.~65, no.~15,
  pp. 4075--4089, 2017.

\bibitem{liu2017one}
C.~Liu and P.~Vaidyanathan, ``One-bit sparse array {DoA} estimation,'' in
  \emph{2017 IEEE International Conference on Acoustics, Speech and Signal
  Processing (ICASSP)}.\hskip 1em plus 0.5em minus 0.4em\relax IEEE, 2017, pp.
  3126--3130.

\bibitem{dabeer2006signal}
O.~Dabeer and A.~Karnik, ``Signal parameter estimation using 1-bit dithered
  quantization,'' \emph{IEEE Transactions on Information Theory}, vol.~52,
  no.~12, pp. 5389--5405, 2006.

\bibitem{gray1993dithered}
R.~Gray and T.~Stockham, ``Dithered quantizers,'' \emph{IEEE Transactions on
  Information Theory}, vol.~39, no.~3, pp. 805--812, 1993.

\bibitem{carbone1997quantitative}
P.~Carbone, ``Quantitative criteria for the design of dither-based quantizing
  systems,'' \emph{IEEE Transactions on Instrumentation and Measurement},
  vol.~46, no.~3, pp. 656--659, 1997.

\bibitem{robinson2019analog}
I.~Robinson, J.~Toplicar, and J.~Heston, ``Analog to digital conversion using
  differential dither,'' May~21 2019, {US} Patent 10,298,256.

\bibitem{ali2020background}
A.~Ali and P.~Gulati, ``Background calibration of reference, {DAC}, and
  quantization non-linearity in {ADCS},'' Jan.~28 2020, {US} Patent 10,547,319.

\bibitem{foucart2019recovering}
S.~Foucart and R.~G~Lynch, ``Recovering low-rank matrices from binary
  measurements,'' \emph{Inverse Problems \& Imaging}, vol.~13, no.~4, 2019.

\bibitem{foucart2019one}
S.~Foucart and L.~Jacques, ``One-bit sensing of low-rank and bisparse
  matrices,'' in \emph{2019 13th International conference on Sampling Theory
  and Applications (SampTA)}.\hskip 1em plus 0.5em minus 0.4em\relax IEEE,
  2019, pp. 1--4.

\bibitem{leventhal2010randomized}
D.~Leventhal and A.~S. Lewis, ``Randomized methods for linear constraints:
  convergence rates and conditioning,'' \emph{Mathematics of Operations
  Research}, vol.~35, no.~3, pp. 641--654, 2010.

\bibitem{ameri2018one}
A.~Ameri, J.~Li, and M.~Soltanalian, ``One-bit radar processing and estimation
  with time-varying sampling thresholds,'' in \emph{2018 IEEE 10th Sensor Array
  and Multichannel Signal Processing Workshop (SAM)}.\hskip 1em plus 0.5em
  minus 0.4em\relax IEEE, 2018, pp. 208--212.

\bibitem{eamaz2023covariance}
A.~Eamaz, F.~Yeganegi, and M.~Soltanalian, ``Covariance recovery for one-bit
  sampled stationary signals with time-varying sampling thresholds,''
  \emph{Signal Processing}, vol. 206, p. 108899, 2023.

\bibitem{candes2015phase}
E.~J. Candes, Y.~C. Eldar, T.~Strohmer, and V.~Voroninski, ``Phase retrieval
  via matrix completion,'' \emph{SIAM review}, vol.~57, no.~2, pp. 225--251,
  2015.

\bibitem{van1996matrix}
C.~Van~Loan and G.~Golub, \emph{Matrix computations}.\hskip 1em plus 0.5em
  minus 0.4em\relax The Johns Hopkins University Press, 1996.

\bibitem{meka2008rank}
R.~Meka, P.~Jain, C.~Caramanis, and I.~Dhillon, ``Rank minimization via online
  learning,'' in \emph{Proceedings of the 25th International Conference on
  Machine learning}, 2008, pp. 656--663.

\bibitem{recht2011null}
B.~Recht, W.~Xu, and B.~Hassibi, ``Null space conditions and thresholds for
  rank minimization,'' \emph{Mathematical programming}, vol. 127, no.~1, pp.
  175--202, 2011.

\bibitem{needell2010randomized}
D.~Needell, ``Randomized {K}aczmarz solver for noisy linear systems,''
  \emph{BIT Numerical Mathematics}, vol.~50, pp. 395--403, 2010.

\bibitem{huang2022linear}
M.~Huang and Y.~Wang, ``Linear convergence of randomized {K}aczmarz method for
  solving complex-valued phaseless equations,'' \emph{SIAM Journal on Imaging
  Sciences}, vol.~15, no.~2, pp. 989--1016, 2022.

\bibitem{plan2014dimension}
Y.~Plan and R.~Vershynin, ``Dimension reduction by random hyperplane
  tessellations,'' \emph{Discrete \& Computational Geometry}, vol.~51, no.~2,
  pp. 438--461, 2014.

\bibitem{carlsson2021lipschitz}
M.~Carlsson, ``{L}ipschitz continuity for isotropic matrix functions,''
  \emph{Linear Algebra and its Applications}, vol. 624, pp. 259--266, 2021.

\bibitem{andersson2016operator}
F.~Andersson, M.~Carlsson, and K.~Perfekt, ``Operator-{L}ipschitz estimates for
  the singular value functional calculus,'' \emph{Proceedings of the American
  Mathematical Society}, vol. 144, no.~5, pp. 1867--1875, 2016.

\end{thebibliography}

\end{document}